\UseRawInputEncoding

\pdfoutput=1

\documentclass[twocolumn,trackchanges]{aastex631}

\usepackage{textcomp}
\usepackage{threeparttable}

\begin{document}

\title{Transformer-based Approach for Accurate Asteroid Spectra taxonomy and albedo estimation }

\correspondingauthor{Y.-J. Tang}
\author{Y.-J. Tang}
\affiliation{School of Physics,\\ Zhejiang University  of Technology,\\ Hangzhou 310023, China}
\affiliation{Collaborative Innovation Center for Bio-Med Physics Information Technology of ZJUT, \\Zhejiang University of Technology, \\Hangzhou 310023, China}

\author{Y.-X. Jiang}
\affiliation{School of Physics,\\ Zhejiang University  of Technology,\\ Hangzhou 310023, China}
\affiliation{Collaborative Innovation Center for Bio-Med Physics Information Technology of ZJUT, \\Zhejiang University of Technology, \\Hangzhou 310023, China}

\author{Y.-X. Feng}
\affiliation{School of Physics,\\ Zhejiang University  of Technology,\\ Hangzhou 310023, China}
\affiliation{Collaborative Innovation Center for Bio-Med Physics Information Technology of ZJUT, \\Zhejiang University of Technology, \\Hangzhou 310023, China}

\author{X.-M. Zhang}
\affiliation{CAS Key Laboratory of Optical Astronomy,\\  National Astronomical Observatories,\\  Chinese Academy of Sciences, Beĳing 100101, China}

\author{X.-J. Jiang}
\affiliation{CAS Key Laboratory of Optical Astronomy,\\  National Astronomical Observatories,\\  Chinese Academy of Sciences, Beĳing 100101, China}




\begin{abstract}

China plans to launch a probe (Tianwen-2) around 2025, mainly for exploring the near-Earth asteroid 2016 HO3 (469219, Kamo'oalewa). The mission involves close-range exploration, landing, and mining operations that require three-dimensional modeling of the asteroid, which requires prior knowledge of its material composition and uniformity. This information is crucial in progressive or ground exploration processes. Our research focuses on high-precision intelligent inversion of complex physical properties of asteroids based on spectral data, providing support for further analysis of asteroid materials, density, and structure. We have developed a platform for asteroid spectral classification, albedo estimation, and composition analysis, which includes three types of neural networks based on Transformer attention mechanism: One for spectral classification, achieving a  four-class classification accuracy of 97.28\% and an eleven-class classification accuracy of 95.69\%; second one for albedo estimation, with an average absolute error of 0.0308 in S-type asteroid albedo estimation, and the third one for composition analysis, with a predicted spectral angular distance of only 0.0340 and a root mean square error of 0.1759 for the abundance of end members. These results indicate that our network can provide high-precision asteroid spectral classification, albedo estimation, and composition analysis results. In addition, we utilized the platform to analyze and provide results for six asteroids.

\end{abstract}



\section{Introduction} \label{sec:intro}

The study of the physical properties of asteroids can reveal the evolution of the Solar System and the origin of life on Earth, provide a basis for evaluating the risk and consequences of near-Earth asteroid impacts, and developing the utilization of asteroid resources.Asteroid 2016 HO3 is the nearest and most stable current Earth quasi-satellite \citep{article1,sktcxb-11-4-lichunlai} and a potential target for future missions\citep{article2}. China plans to launch a probe (Tianwen-2) to detect 2016 HO3 around 2025 \citep{article3}. The mission plans to conduct orbiting, sampling (touching, hovering, attachment) and samples return in one mission\citep{article4}. However, due to its faintness, our understanding of 2016 HO3 remains limited. Apart from the exploration of 2016 HO3, the Tianwen-2 mission also encompasses the investigation of the orbit of the main-belt comet 311P. It aims to achieve the detection of two targets in one launch, realizing three modes of exploration \citep{article5}.

Spectral analysis and taxonomical classification of asteroids is an important tool for conducting asteroid research. By analyzing the spectral data, it is possible to categorize asteroids  into distinct taxonomical classes estimated values of albedo can be obtained based on the classes, and preliminarily determine their composition and mineral components based on the reflection and absorption characteristics of the asteroids. This is crucial for both the planning and execution of the mission.

Multiple asteroid classification systems have been formed. In Chapman et al. (\citeyear{article6,chapman1975surface}) pioneered a classification framework for asteroids based on color, reflectance, and spectral features, dividing asteroids into different categories. These properties are believed to correspond to the composition of asteroid surface materials \citep{bowell1978taxonomy}. For example, the initial carbonaceous C-type, siliceous S-type, and metallic M-type are evident in different observations. Subsequently, with the accrual of extensive spectral datasets about asteroids, the taxonomy has undergone successive refinements to encompass finer delineations. Presently, the prevalent taxonomical methodologies include the Tholen taxonomy and the Bus-DeMeo taxonomy. The Tholen taxonomy, introduced by Tholen et al. (\citeyear{article7,article8}) based on observations from the Eight-Color Asteroid Survey, delineates asteroids into 14 distinct types. In 2002, Bus, Binzel, et al. (\citeyear{article9,article10})  expanded the taxonomic scope to 26 classes utilizing data from the Small Main-Belt Asteroid Spectroscopic Survey II (SMASS II), forming the Bus \& Binzel taxonomy. However, acknowledging the limitation of considering solely visible light spectra, DeMeo et al. (\citeyear{article11}) incorporated infrared data to refine the Bus \& Binzel taxonomy, yielding the Bus-DeMeo taxonomy. This updated classification system categorizes asteroids into four primary classes: S, C, X, and others, further subdividing these classes into 24 subclasses.
   
   The spectral class of an asteroid is closely related to its composition \citep{demeo2015compositional}. The overall shape and absorption features of the reflectance spectra of asteroids reflect the composition and minerals they contain. Asteroids of the same taxonomy often have a similar composition. The spectral types of many asteroids may have the same mineralogy, but some changes may not be the result of mineralogy, but rather differences caused by surface texture, degree of space weathering, other unknown influences of the space environment, and variations or problems in the observation techniques used \citep{article9,article10}. There are also some asteroids with different mineralogy but the same spectral taxonomy. For example, the X-type asteroids are several populations with spectra that are very similar, but have different compositions. Asteroid classification is based on the observed spectra of asteroids (usually hundreds of asteroids in a given number of colors or continuous spectra) and their spectral characteristics. Therefore, taxonomy can tell us whether the observed spectra of asteroids contain absorption bands of specific minerals, but they contain little information about the quantitative mineral abundance or the mineral chemical composition of asteroids. Taxonomy can serve as a reference for asteroid mineralogy. In many cases, this reference is quite accurate.
   
   S-type asteroids encompass a variety of compositional subtypes, the silicate composition on its surface ranging from pure olivine (dunites) to olivine-pyroxene mixtures, to orthopyroxene or orthopyroxene-feldspar mixtures (basalt). Moreover, spectral slopes indicate the presence of Fe-Ni metal \citep{article12,article13}. Chondrite-like meteorites of C-type asteroids are rich in volatile carbonaceous chondrules, including CI, CM, C2, and CR carbonaceous chondrites. These meteorites feature abundant hydrated silicates and magnetite. Minor minerals include calcite, dolomite, olivine, a suite of hydrated clays, and varying amounts of organic material \citep{article14}. X-type asteroid spectra yield little information beyond a moderate to strong spectral slope. Their meteoritic counterparts are iron meteorites, believed to be composed primarily of Fe-Ni metal, orthopyroxene, magnesium olivine, and feldspar. They are primarily found among the innermost members of the asteroid belt, particularly within the Hungaria asteroid group \citep{article15,article16}. However, taxonomies can tell us whether the spectrum of the observed asteroid contains absorption bands of specific minerals, but they contain little information about the quantitative mineral abundances or mineral chemical compositions of the asteroid. Therefore, upon classifying asteroid spectra, we need to further employ spectral deconvolution methods to obtain quantitative \textbf{indicators} of their minerals and chemical compositions, thereby determining the types and distribution of resources present on the asteroid surfaces, such as water, metallic minerals, rare elements, etc. Establishing the distribution of these resources is of paramount significance for the future development and utilization of space resources.
   
   In recent years, machine learning methods have been widely used in the taxonomy and compositional study of asteroids.  For example, Bus \& Binzel (\citeyear{article10}) use the PCA method \citep{smith2002tutorial} to reduce the dimensionality of the data and classify asteroids. Huang et al.  (\citeyear{article17}) employed a random forest algorithm to conduct an eight-class classification of the observations from the Sloan Digital Sky Survey Moving Object Catalogue (SDSS MOC) . Klimczak et al. (\citeyear{article18}) confirmed that machine learning methods such as naive Bayes, support vector machine (SVM), gradient boosting, and multilayer networks can reproduce that taxonomic classification at a high rate of over 81\% balanced accuracy for types.    Deep learning, as an emerging subfield of machine learning, has also made significant progress in related fields. Penttilä et al. (\citeyear{article19}) explored the performance of neural networks in automatic classification of asteroids and constructed a network that can robustly identify the classification taxonomy of asteroids, with an accuracy rate of 90.2\% .  Luo et al. (\citeyear{article21}) designed a classification system using a neural network (NN) algorithm with an accuracy exceeding 92\%, planned for analyzing China Space Station Telescope (CSST) asteroid spectra.  Korda et al. (\citeyear{article22,article23}) designed a convolutional neural network (CNN) with two hidden layers to analyze asteroid spectra, perform classification, and analyze surface components, testing the data on (433) Eros and (25143) Itokawa asteroids . Mahlke et al.(\citeyear{article20}) established a classification method consisting of 3 major classes and 17 subclasses. They reintroduced albedo into observational classification, akin to Tholen, and proposed a mixture of common factor analysers (MCFA) mixed-model approach for asteroid data classification. They provided their classification of 17 types of asteroids and the albedo range for each type of asteroid.

   After classifying the spectra of asteroids, their spectra can be analyzed by unmixing. By decomposing the composition and proportion of each material spectrum in the asteroid spectrum, the main material composition of the asteroid can be analyzed. The research on spectral unmixing has developed a large number of theories and methods around the mathematical problems of decomposition. According to the interaction between light and matter, spectral unmixing methods can be divided into two categories: linear and nonlinear unmixing; And further corresponding to the different focus areas of endmember extraction, abundance estimation, and unsupervised unmixing in the unmixing process, there are also matching method branches.

   In the mission of planetary remote sensing exploration, \textbf{obtaining} the types, distribution, and physicochemical properties of minerals on the surface of planets helps to understand the history and spatial evolution of planets, and is also of great significance for possible future space exploration activities such as extraterrestrial migration and resource exploitation. Parente et al. (\citeyear{parente2011robust}) used the nonnegative matrix factorization (NMF) method with endmember smoothing constraints to extract local endmembers from the clustered sub region images in the analysis of hyperspectral remote sensing images for Mars exploration. After screening, they obtained a candidate endmember set that approximates the actual reference. Lin Honglei et al. (\citeyear{LINGHonglei2019}) successfully achieved the inversion of mineral abundance and the drawing of particle size distribution curves in the spectral data of Yutu rover by combining the single scattering albedo conversion results of the collected data with the radiative transfer Hapke model and the sparse unmixing algorithm SUnSAL. Yin et al. (\citeyear{yin2019automatic}) used sparse representation and least-squares regression method considering endmember beams to unmix the spectral images of Chang'e-1 and accurately estimat the abundance distribution of lunar surface minerals. Ling-Zhi et al. (\citeyear{ling2014spectral}) used the modified Gaussian model (MGM) for spectral decomposition of mixed minerals to obtain mineral endmembers, and used spectral angle mapper (SAM) to create a distribution map of the main minerals in the Maunder impact crater. Afterwards, Ling-Zhi et al. (\citeyear{LingzhiSun2016}) used the Lunar Mineral Plotter (Moon mineralogy mapper) to obtain fresh hyperspectral data of impact craters near the landing area of Chang'e-3. They used the Hapke radiative transfer model and modified the Gaussian model (MGM) to jointly analyze and quantitatively invert the ferromagnesian minerals in the young lunar mare basalt. Research has shown that the mineral composition of basalt in this region is mainly composed of clinopyroxene minerals, with a high proportion of olivine. Zhang Qi et al. (\citeyear{ZHANGQi2016}) proposed a method for linearly decomposing hyperspectral remote sensing data on the lunar surface to obtain mineral content. Firstly, the Hapke radiative transfer model is used to convert the nonlinear mixed reflectance spectra of five minerals in the Relab spectral library into a linear mixed single albedo. Then, mixed pixels are randomly generated based on the proportion. Finally, based on the fully constrained linear spectral decomposition method, a statistical relationship model is established between the decomposition content and the true content of the five minerals mentioned above.

   Normally, the albedo of asteroids is obtained by fitting theoretical thermal models with actual observational data. For example, assuming that an asteroid has a spherical shape, with a smooth surface and an observed phase angle of  $0^\circ$, the Standard Thermal Model (STM) is applicable to main belt asteroids \citep{lebofsky1986refined}. For asteroids with fast rotation speed and large thermal inertia, use the Fast Rotation Model (FRM) can be used to derive their effective diameter and albedo \citep{lebofsky1989radiometry}. For general near-Earth asteroids, a near-Earth asteroid thermal model (NEATM) was established to derive the relevant equivalent diameter and albedo \citep{harris1998thermal}. In addition, asteroid taxonomy reflect the different chemical and physical properties of asteroids, which are related to their albedo. Therefore, the albedo can also provide a rough estimate based on the asteroid taxonomy. 
   
   Although deep learning methods have made some progress in the classification and composition analysis of asteroid spectra, the structure of these artificial intelligence networks is still relatively simple, consisting only of a few convolutional layers or fully connected layers, resulting in low accuracy in the final results. Given that an erroneous classification of asteroid categories and mineral compositions during the Tianwen-2 mission could result in significant losses, we aim to introduce more advanced artificial intelligence architectures and modules from the field of deep learning to enhance the accuracy of predictions further.
      
   In recent years, the Transformer network has achieved tremendous success in various artificial intelligence fields due to its powerful feature extraction ability \citep{article24,article25}. As one of the key components of the Transformer, the attention mechanism enables the correlation and interaction of information from different positions in the input sequence, thereby enhancing the model's representational and generalization abilities. This article establishes a platform for asteroid spectral classification, albedo estimation, and composition analysis based on a multi head self attention module, and for the first time applies this network to asteroid data analysis tasks. For the input asteroid spectra, the asteroid spectral classification network (ASC-Net) spectral classification network we constructed determines the asteroid categories, and using the asteroid albedo estimation network (AAE-Net) to invert the estimated value of the albedo based on spectral and classes information. The asteroid's quantitative compositional abundance is then determined by the spectral unmixing network auto encoder transformer network (AE-trans).

   This paper is arranged as follows. Section 2 describes the preparation of the training and test data to be used with our networks. Section 3 introduces the construction of our spectral classification, albedo estimation and compositional analysis platform. Section 4 introduced the training process and accuracy evaluation of ASC-Net, AE-trans, and AAE-Net, and demonstrated the practical application of the constructed platform, as well as some analysis results of asteroid spectra. Section 5 gives a summary of the work.

\section{Data Preparation} \label{sec:style}
We constructed, three neural networks, ASC-Net, AE-trans, and AAE-Net used for asteroid spectral classification, albedo estimation, and composition analysis. Therefore, we need to prepare three different datasets for training and testing the three neural networks separately.

\subsection{Asteroid Spectral Taxonomy Dataset} \label{subsec:tables}

\begin{figure}[h]
    \centering
    \includegraphics[width=8.5cm]{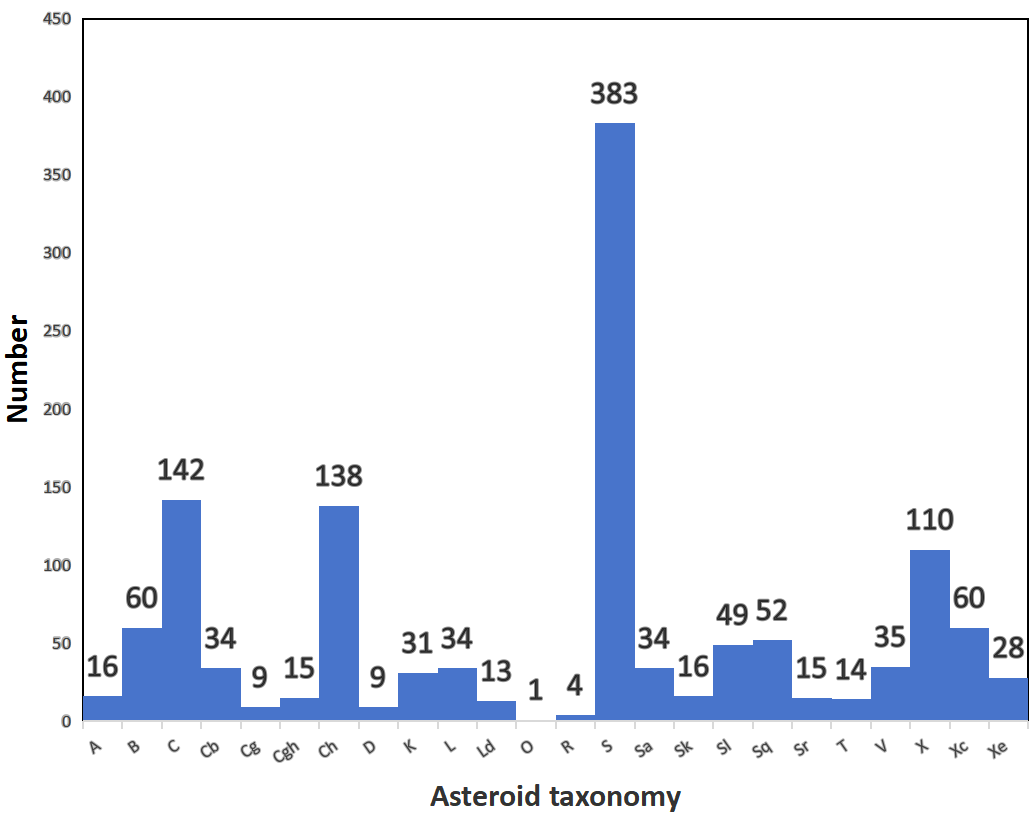}
    \begin{center}
    \caption{Spectral counts of various types of asteroids: The horizontal axis represents the taxonomy of asteroids, and the vertical axis represents the number of spectra of such asteroids. }
     \label{fig:1}
    \end{center}
\end{figure}

To train and test the ASC-Net neural network for asteroid spectral classification, we constructed an asteroid spectral taxonomy dataset (ASTD). The spectral data in the dataset was sourced from the SMASS II database and consisted of 1341 asteroid spectral samples with a sampling interval of 0.01 {\textmu}m and a wavelength range of 0.44 {\textmu}m to 0.92 {\textmu}m. We assign taxonomy labels to each spectral data using the Bus \& Binzel classification method \citep{article9,article10}. The number of asteroid spectral samples in each taxonomy is shown in Figure \ref{fig:1}. 

Firstly, we removed the O and R categories with insufficient spectral samples. In the remaining 23 categories, there is still a serious issue of data imbalance. Direct use of this dataset for training and testing may lead to overfitting of the algorithm on the categories with more samples and underfitting on those with fewer samples. Therefore, we have decided to expand our dataset by referring to the method proposed by Penttilä et al. (\citeyear{article19}). The method involves performing PCA (Principal composition analysis) on spectral data, then adding random noise in the principal component space, and finally obtaining new spectral data through inverse PCA transformation. 

  Our purpose of spectral classification is for the analysis of material components, and the existing research on the material composition of asteroids is mainly based on the Bus-DeMeo four major taxonomy classes, there is relatively little research on the material composition of each subclass. Therefore, our platform is based on the four major classes in actual operation. The classification to subclasses of asteroids will be considered in future work. The comparison between the four classes of Bus-DeMeo taxonomy and the Bus \& Binzel taxonomy method is shown in Table  \ref{tab:1}.

     \begin{table}[h]
   \caption{Comparison of the Bus-DeMeo taxonomy and the Bus\&Binzel taxonomy}
    \centering
    \begin{tabular}{cc}
        \hline
       Bus-DeMeo\hspace{0.5em}taxonomy       &  Bus \& Binzel\hspace{0.5em}taxonomy \\
            \noalign{\smallskip}
            \hline
            \noalign{\smallskip}
            S& S, Sa, Sk, Sl, Sq, Sr     \\
            C & B, C, Cb, Cg, Cgh, Ch            \\
            X          & X, Xc, Xe, Xk \\
            Other     & A, D, K, L, Ld, O, R, Q, T, V   \\
            \noalign{\smallskip}
            \hline
    \end{tabular}
\label{tab:1}
\end{table}

\begin{figure}[h]
    \centering
    \includegraphics[width=9cm]{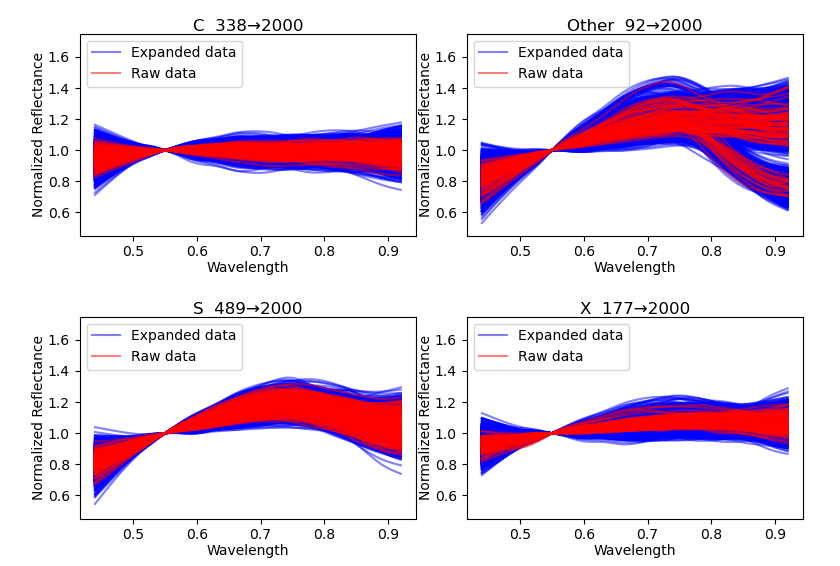}
    \begin{center}
    \caption{Four types of spectral data augmentation display, with original spectral data (red) and expanded spectral data (blue). }
     \label{fig:ASTD_exp}
    \end{center}
\end{figure}

For the three major categories S, C, and X, the surface material compositions are quite similar, while asteroids that cannot be classified into these three categories are grouped into the Other type.  We labeled 1341 spectral data in the dataset as four class data. Subsequently, 60 samples were selected from each type, resulting in a total of 240 samples as the test dataset. Afterwards, the remaining 1101 samples were expanded with 2000 samples per class, resulting in a total of 8000 samples as the training dataset. The test dataset does not participate in expansion, ensuring independence from the training dataset. The dataset before and after expansion is shown in Figure \ref{fig:ASTD_exp}.  

\subsection{Asteroid Composition Analysis Dataset} \label{subsec:tables}

\begin{figure*}[ht]
    \centering
    \includegraphics[width=18cm]{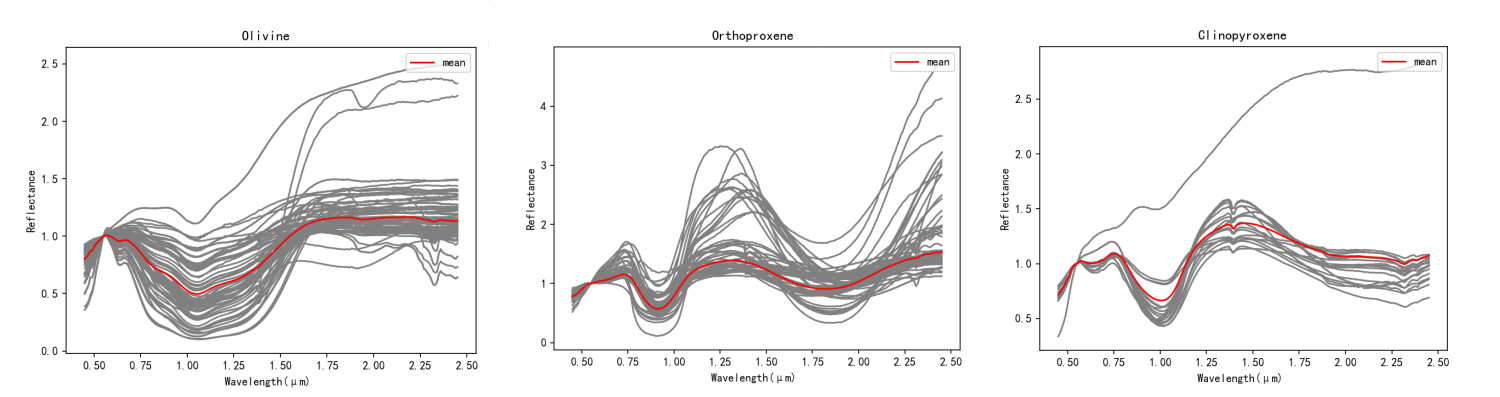}
    \begin{center}
    \caption{End-member spectra of various categories of  S-type asteroids. (left): Olivine.(center): Orthopyroxene. (right): Clinopyroxene. }
    \label{fig:2}
    \end{center}
\end{figure*}

The target asteroid of Tianwen-2, 2016HO3, belongs to the S-type asteroids and is rich in silicate minerals such as olivine and pyroxene. We constructed an Asteroid Composition Analysis Dataset (ACAD) for training and testing the S-type asteroid spectral unmixing network AE-trans. In ACAD, We selected olivine, orthopyroxene, and clinopyroxene as endmembers for compositional analysis of S-type asteroids. The spectral data of these materials are all sourced from the RELAB database\footnote{\href{https://sites.brown.edu/relab/relab-spectral-database}{https://sites.brown.edu/relab/relab-spectral-database}}. These spectrum covering the wavelength range 0.45 {\textmu}m to 2.45 {\textmu}m, with a sampling interval of 0.01 {\textmu}m . The data we have selected is shown in Figure \ref{fig:2} .

Regarding the Spectral unmixing dataset (ACAD), it contains two simulated datasets, both generated using the same method. Two 3 × 2500  content matrix A1 and A2 are randomly generated and combined with a 201 × 3 spectral matrix M composed of three material spectra. Then, they are normalized at 0.55 \textmu m to obtain two 201 × 2500 simulated data matrices, Simulated1 and Simulated2. This approach allows us to explicitly determine the abundance of each end-member in every generated spectrum. We use Simulated1 to train the network and use the simulated dataset Simulated2 and the composition dataset AS used by Korda D et al.(\citeyear{article22}) to test the network. 

\subsection{Asteroid Albedo Estimation Dataset} \label{subsec:tables}
Based on the asteroids included in the collected spectral dataset, we downloaded the albedo related information of the corresponding asteroids from the SSODnet\footnote{\href{https://sites.brown.edu/relab/relab-spectral-database}{https://ssp.imcce.fr/webservices/ssodnet}}, combine it with the spectral dataset to obtain the Asteroid Albedo Estimation Dataset (AAED).

The dataset we collected includes spectral data of 494 asteroids ranging from 0.45{\textmu}m to 2.45{\textmu}m and their albedo data. The albedo distribution of three different types of asteroids in the dataset is shown in the following  Table \ref{tab:AAEDDATA}
   \begin{table}[h]
      \caption{Mean and standard deviation of albedo of different types of asteroids in AAED}
        \centering
         \begin{tabular}{ccc}
            \hline
            \noalign{\smallskip}
            Type & Number     & Albedo\\
            \noalign{\smallskip}
            \hline
            \noalign{\smallskip}
            S& 220 &$0.2298\pm0.0697$    \\
            C & 64   & $0.0606\pm0.0592$        \\
            X          & 39 &$0.2187\pm0.2132$\\
            \noalign{\smallskip}
            \hline
        \end{tabular}
        \label{tab:AAEDDATA}
   \end{table}

\section{Neural Networks} \label{sec:floats}

On our asteroid spectral classification and composition analysis platform, we have carefully constructed three neural networks: ASC-Net for asteroid spectral classification, AAE-Net for albedo estimation, and AE-trans for asteroid composition analysis. 
\subsection{Attention mechanism}
\label{subsec:tables}

\begin{figure}[h]
    \centering
    \includegraphics[width=8cm]{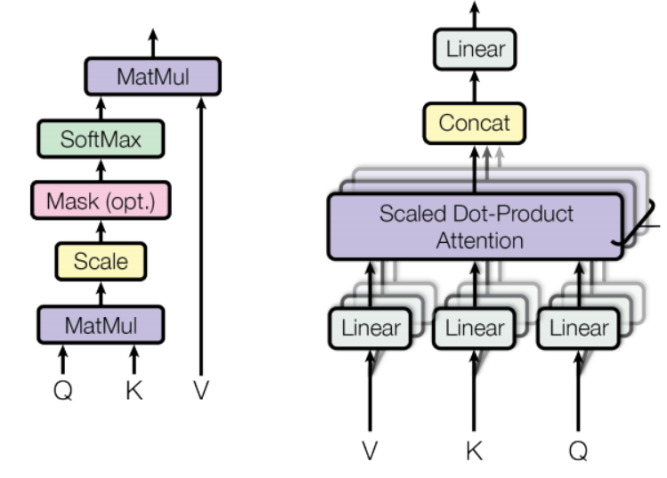}
    \begin{center}
    \caption{The structure of the attention mechanism: self-attention mechanism  (left); multi-head self-attention mechanism  (right)  \citep{yang2019scrdet} }
    \label{fig:Attention}
    \end{center}
\end{figure}

As one of the most advanced and widely used modules in the field of deep learning, the attention mechanism has a powerful feature extraction capabilities and can associate and interact with information from different positions in the input sequence. \citep{article25} Neural networks that incorporate self-attention mechanism have stronger representation and generalization capabilities, which means that the network can achieve higher classification and composition analysis accuracy on new asteroid \textbf{spectrum}. The module structure is shown in the Figure \ref{fig:Attention}

In the computation process of self-attention mechanism, each value of the input sequence is first transformed into three different vectors through linear mapping: query vector $q$ , key vector $k$ , and value vector $v$. Subsequently, the similarity $S$ between the query vector and the keyword vector is calculated by dot product:
\begin{equation}
\textbf{$S$}=\textbf{$q$} \cdot\textbf{$k$}_{}^{T}\label{eq:1}
\end{equation}
Subsequently, the similarity \textbf{$S$} is normalized, where 
\textbf{$d_k$}
is the dimensionality of the key vector
\textbf{$k$}:
\begin{equation}
S_n=\frac{S}{\sqrt{d_k}}\label{eq:2}
\end{equation}
After obtaining the normalized similarity \textbf{$S_n$}, the attention weight {$P$} is further computed by applying the Softmax function.
\begin{equation}
P=Softmax(S_n)
\end{equation}
By utilizing the attention weight to perform weighted summation over the value vectors \textbf{$v$}, the output representation of each position can be obtained.
\begin{equation}
\textbf{$Z$}=\textbf{$v$}\cdot\textbf{$P$}
\end{equation}
Finally, concatenating all the output representations of positions together yields the final output of the self-attention layer.

In practical use, when inputting a set of data into a model, we often hope that the model can simultaneously pay attention to key information at different positions in the entire set of data. However, using a single head self attention mechanism can lead to the model excessively focusing its attention on its own position when encoding information about the current location. Therefore, it is necessary to solve this problem by combining multi-head self- attention mechanism.

The multi-head self-attention mechanism sets \textit{n} sets of \textbf{$k$}, \textbf{$q$}, and \textbf{$v$} values. After training each set of values through random initialization, the input vector is projected onto different feature subspaces and various self-attention layer outputs are calculated. Finally, the attention calculation results of all heads are concatenated and linearly mapped again to obtain the final multi head self attention output \citep{han2022survey}.
In addition, the advantage of multi-head attention mechanism is that the model can learn different behaviors based on the same attention mechanism, and then combine different behaviors as knowledge.

\subsection{Asteroid Spectral Classification Network}
\label{subsec:tables}
\begin{figure}[!h]
    \centering
    \includegraphics[width=9cm]{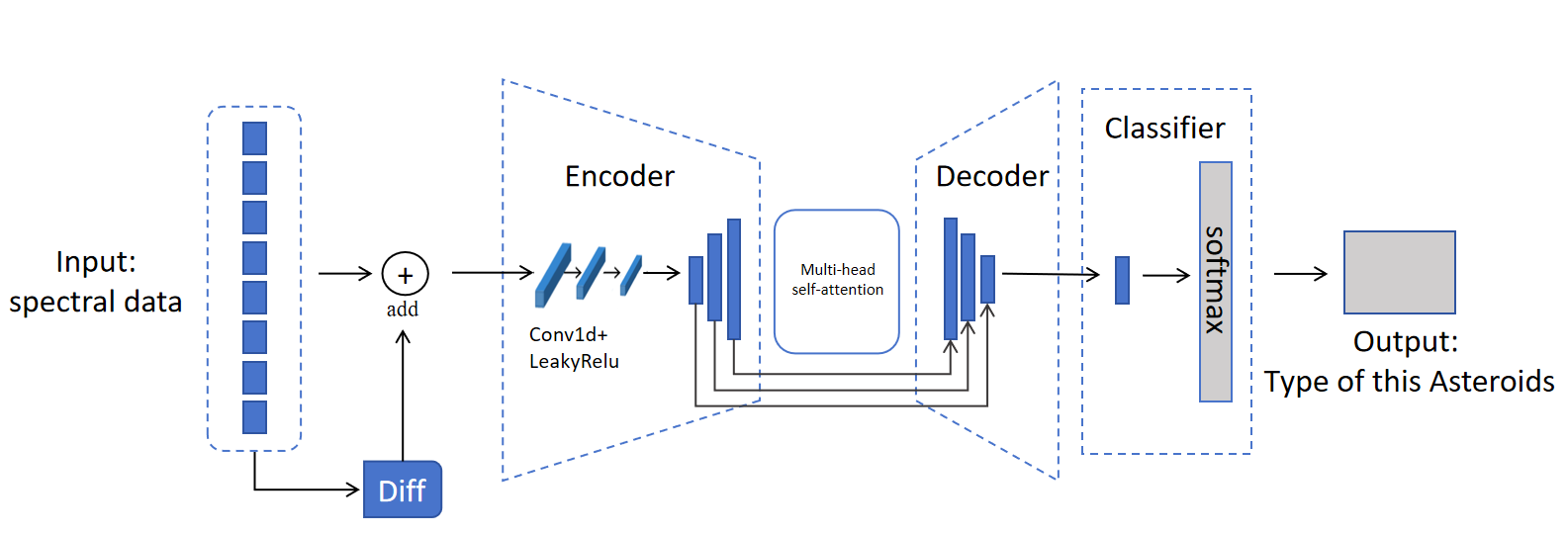}
    \begin{center}
    \caption{Network structure of ASC-Net: The fusion of spectral data and differentiated extended data is input into an encoder composed of three convolutional layers and three fully connected layers, and then input into a decoder composed of fully connected layers through a multi head attention mechanism module. Three skip connections are used to improve feature extraction ability, and then the categories are output through a softmax layer.}
    \label{fig:4}
    \end{center}

\end{figure}

The overall architecture of our intelligent classification network Asteroid Spectral Classification Network (ASC-Net) based on attention mechanism is illustrated in Figure  \ref{fig:4}. The network is designed as an encoder-decoder structure.

In the process of classifying asteroid spectra, the normalized spectra of different asteroid classes, including the positions of spectral peaks and valleys, as well as the slope of the curves, serve as crucial criteria for spectral classification. Therefore, in the data input layer, we incorporate a differential module, referred to as Diff, to expand the input information. The differential information is concatenated with the original spectra as different feature channels and fed into the encoder. The differential module highlights information such as the slope and peak-valley positions of the input spectra, which aids in feature extraction by the network, thereby enhancing classification accuracy and robustness. The calculation method of the differential module is presented in Equation \ref{eq:6}.
\begin{equation}
{diff} _i=r_i-r_{i-1}\label{eq:6}
\end{equation}
In the above equation, ${diff} _i$ represents the difference between the reflectance $r_i$ of the \textit{i}-th band and the reflectance $r_{i-1}$ of the \textit{i}-1-th band in the spectrum.

The encoder of the ASC-Net network consists of three one-dimensional convolutional layers, with kernel sizes of 3, 7, and 3, and stride values set to 1, the activation function is Leaky ReLu. During the network operation, the encoder encodes the input information and inputs the encoded feature information into the decoder. 
The encoder backend of the classification network consists of three fully connected layers, with 128, 256, and 512 neurons respectively. It upsamples the input feature information and inputs the upsampled feature vectors into a multi head self attention mechanism. The multi-head attention mechanism extracts contextual information and captures potential correlations between different positions in the sequence. 
Then, classification counting is performed through the 512 and 256 fully connected layers in the decoder. In order to improve the feature extraction of small sample datasets, the first three fully connected layers and the last three fully connected layers are connected by skip connections, and the softmax classification layer is used to output taxonomy labels and corresponding taxonomy probabilities. 

The activation function chosen for the ASC-Net network is Leaky ReLU. Leaky ReLU is a commonly used activation function in neural networks. It is an improved version of the traditional ReLU function, allowing for a small non-zero gradient when the input is negative.

The main advantage of Leaky ReLU over traditional ReLU is that it prevents dead neurons, which are neurons that always output zero and do not contribute to the learning process. By allowing a small gradient for negative inputs, Leaky ReLU ensures that even neurons that are not active can still update their weights and learn from the data. Additionally, Leaky ReLU addresses the "dying ReLU" problem, which can occur when a large portion of the network's neurons become inactive and stop learning altogether. This can happen if the learning rate is set too high or if the neurons receive large negative inputs, causing them to always output zero with traditional ReLU. With Leaky ReLU, even these neurons can contribute to the learning process, improving the overall performance of the network.

$L1$ Loss is chosen as the loss function. During the network training process, we set a dropout probability of 0.2. Dropout is a commonly used regularization method for neural networks, which can effectively avoid overfitting. During training, Dropout randomly deletes a certain proportion of neurons, so that only a part of the neurons participate in the calculation each time, thereby forcing different neurons to learn different features, making the network more generalized.
The parameters of ASC-Net are shown in Table \ref{tab:3}.
   \begin{table}[h]
      \caption{Parameters of ASC-Net}
         \centering
         \begin{tabular}{cc}
            \hline
            \noalign{\smallskip}
            Parameters       & Value\\
            \noalign{\smallskip}
            \hline
            \noalign{\smallskip}
            Optimizer& Adam     \\
            Loss\hspace{0.5em}function & L1\hspace{0.5em}Loss            \\
            Activation\hspace{0.5em}function         & Leaky\hspace{0.5em}ReLU \\
            Learning\hspace{0.5em}rate     & 0.0001 \\
            Dropout&0.2\\
            Epochs&120\\
            Batch\hspace{0.5em}size&128\\
            \noalign{\smallskip}
            \hline
            \label{tab:3}
         \end{tabular}
   \end{table}

\subsection{AE-trans composition analysis Network}
\label{subsec:tables}

Because the computational complexity of unmixing is relatively high compared to albedo and classification, we referred to some spectral unmixing networks for modification specifically for unmixing.

\begin{figure}[h]
    \centering
    \includegraphics[width=8cm]{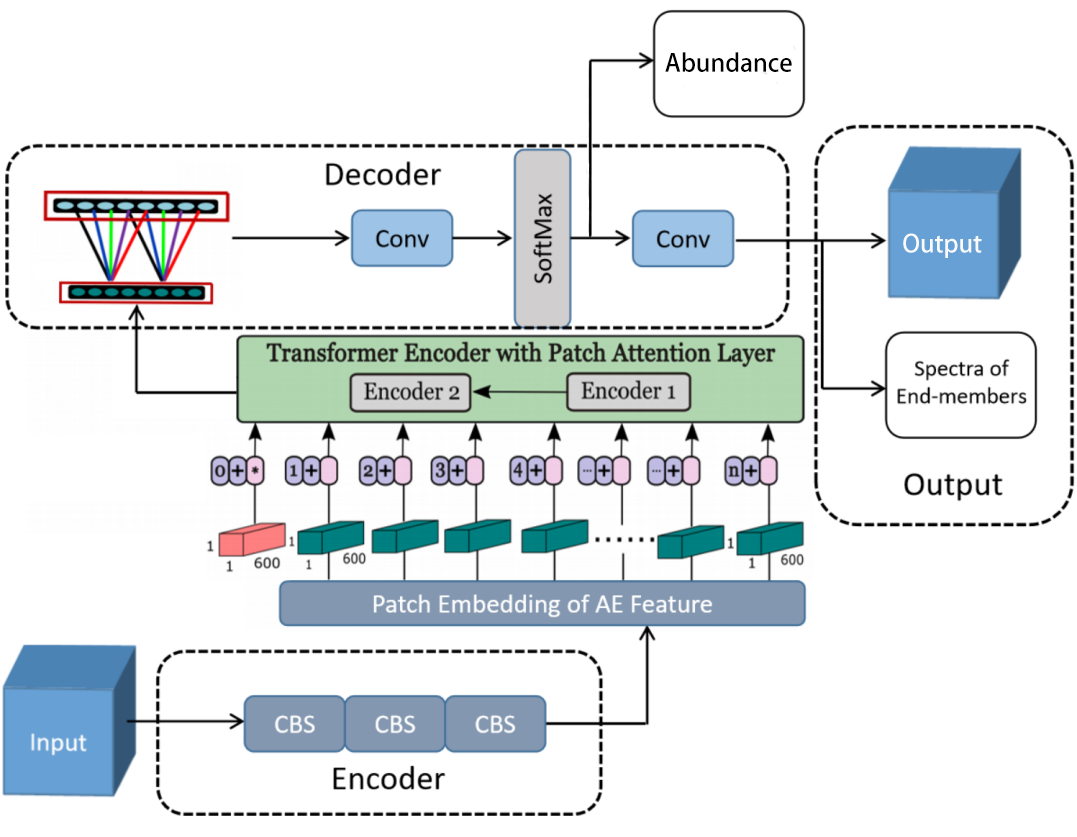}
    \begin{center}
    \caption{Network structure of AE-trans: Input spectral data, pass through three convolutional layers (CNN encoder), and represent discriminative features with fewer channels. The output of the CNN encoder is decomposed into patches. These patches are reshaped into vectors and passed through a transformer encoder consisting of multi head attention layers and MLP layers. Amplify and reshape the output of the transformer encoder to match the size of the abundance map, and use convolutional layers to reduce noise. Further use the softmax activation function to obtain the final abundance map. Finally, the decoder utilizes a single convolutional layer with weights as enddata to reconstruct the represented hyperspectral image Output.}
    \label{fig:5}
    \end{center}
\end{figure}

In the field of spectral unmixing, autoencoder (AE) is a commonly used deep learning method. When performing spectral unmixing, the structure of AE is relatively simple, while also having strong robustness and generalization ability, and exhibits stable performance when dealing with different types of data. We combined AE with Transformer to construct the AE-trans network\citep{ghosh2022hyperspectral}, and improved it for the task of analyzing the composition of asteroids. The structure of AE-trans is shown in Figure \ref{fig:5}.

In the AE-trans network, the encoder is composed of three stacked CBS modules (Conv + Batch normalization + the SiLU). The computation of SiLU activation function is shown in Equation \ref{eq:9}.
\begin{equation}
SiLU=x\cdot Sigmoid(x)\label{eq:9}
\end{equation}

Spectral data is input into three convolutional layers, encoded by a convolutional autoencoder to represent discriminative features with fewer channels and output. Then, a transformer encoder composed of multi-head attention layers and MLP layers is used to capture the correlation between encoded data. Afterward, the output of the transformer encoder is amplified and reshaped to match the size of the abundance map, and convolutional layers are used to reduce noise. Further, the softmax activation function is used to obtain the final abundance map. Finally, a convolutional decoder is used to reconstruct the data, where the weights of the convolutional layer are the \textbf{endmembers}. Due to the characteristics of asteroid spectral data, the convolution kernels used in AE-Trans autoencoders are all 1x1, and the original position and classification encoding parts in the Transformer network have been removed. AE-trans will ultimately output the abundance of different end members in the input spectrum, their corresponding spectra, and reconstructed data.

In order to train the proposed model, two different combinations of losses were used: reconstruction error (RE) loss and spectral angle distance (SAD) loss:
\begin{equation}
L_{RE} \left ( \textbf{\textit{I}},\hat\textbf{\textit{I}}  \right ) =\frac{1}{H\cdot W} \sum_{i=1}^{H} \sum_{j=1}^{W} (\hat\textbf{\textit{I}} _{ij}-\textbf{\textit{I}}_{ij})^{2} \label{eq:re}
\end{equation}
\begin{equation}
L_{SAD} \left ( \textbf{\textit{I}},\hat\textbf{\textit{I}}  \right ) =\frac{1}{R} \sum_{i=1}^{R} arccos 
  \left ( \frac{<\textbf{\textit{I}}_{(i)},{{\hat\textbf{\textit{I}} }}_{(i)}>}{||\textbf{\textit{I}}_{(i)}||_2,||{{\hat\textbf{\textit{I}} }}_{(i)}||_2}\right )\label{eq:SAD}
\end{equation}

In the formula, $\textbf{\textit{I}}$ represents the input spectrum, and $\hat\textbf{\textit{I}}$ represents the output reconstructed spectrum. $R$ represents the number of endmembers, $H$ and $W$ represent the number of input spectral bands, and the number of input spectra.

The RE loss is calculated by the mean squared error (MSE) objective function, which helps the encoder to only learn the basic features of the input HSI and discard unnecessary details. SAD loss is a scale-invariant objective function. MSE distinguishes the final members based on their absolute size, which is not advisable in the case of HSI separation. Including SAD loss helps to counteract the drawback of the MSE objective function and makes the entire model converge faster. The total loss is calculated as the weighted sum of these two losses:
\begin{equation}
L_{loss} =\beta L_{RE}+\gamma  L_{SAD}
\end{equation}
 with regularization parameters $\beta$ and $\gamma$.

The parameters of AE-trans are shown in Table \ref{tab:4}.
   \begin{table}[h]
      \caption{Parameters of AE-trans}
        \centering
         \begin{tabular}{cc}
            \hline
            \noalign{\smallskip}
            Parameters       & Value\\
            \noalign{\smallskip}
            \hline
            \noalign{\smallskip}
            Optimizer& SGD     \\
            Loss\hspace{0.5em}function & SID\hspace{0.5em}Loss            \\
            Activation\hspace{0.5em}function         & SiLU \\
            Learning\hspace{0.5em}rate     & 0.000001 \\
            Dropout&0.2\\
            Epochs&200\\
            Batch\hspace{0.5em}size&128\\
            \noalign{\smallskip}
            \hline
            \label{tab:4}
         \end{tabular}
     
   \end{table}

\subsection{Asteroid Albedo Estimation Network}
\label{subsec:tables}
\begin{figure}[!h]
    \centering
    \includegraphics[width=9cm]{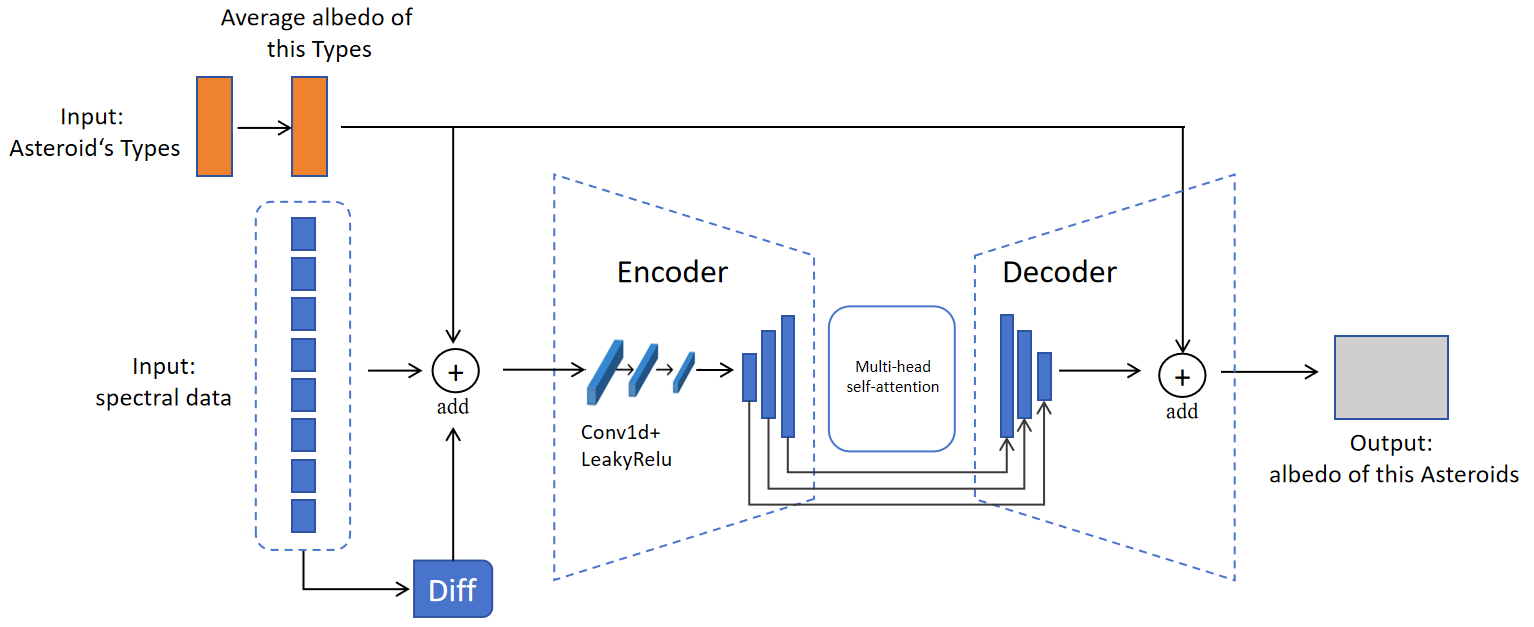}
    \begin{center}
    \caption{Network structure of AAE-net: The fusion of spectral data, differential extension data, and category information is input into an encoder consisting of three convolutional layers and three fully connected layers, and then input into a decoder composed of fully connected layers through a multi head attention mechanism module. The category information is then fused and the albedo is output through the fully connected layers }
    \label{fig:4-1}
    \end{center}

\end{figure}

The overall architecture of the Asteroid Albedo Estimation Network (AAE-Net) based on the attention mechanism intelligent regression network is shown in Figure \ref{fig:4-1}. The network is also designed as an encoder-decoder structure.

We input the average albedo of asteroids in this class as the classified category information into the albedo estimation network.  Based on visible-light and near-infrared spectral bands for albedo estimation, it is also necessary to extract the changes and features of each spectral band as much as possible. Use the differential module Diff to expand the input information and input the differential information and the original spectrum, after adding category information, are used as different feature channels input into the subsequent encoder. The encoder consists of three one-dimensional convolutional layers with an activation function of \textbf{Leaky ReLu}. The encoder encodes the input information and inputs the encoded feature information into the decoder. The decoder part of the albedo estimation network adds feature fusion with category information, and finally directly connected to a fully connected layer with layer 1 to output the albedo.

\section{Results and Discussion} \label{sec:cite}
\subsection{Classification Accuracy of ASC-Net }
\label{subsec:tables}
\subsubsection{Test on ASTD}

Using the ASTD dataset, we conduct training for ASC-Net and test the classification accuracy using 1005 real spectra from the SMASS II database. We evaluate the classification accuracy of the network using two metrics: Accuracy (Acc) and Average Precision (AP). Accuracy refers to the total number of correctly classified instances divided by the total number of instances, and is a commonly used evaluation metric for classification networks. However, due to the imbalance in the number of each taxonomy in the test dataset, we have added average accuracy as the evaluation metric, which calculates the average accuracy for each taxonomy separately. The computation of Accuracy and Average Precision is illustrated in Equations \ref{eq:13} and \ref{eq:14} respectively.
\begin{equation}
Acc=\frac{T}{T+F}\label{eq:13}
\end{equation}
\begin{equation}
AP=\frac{1}{C}\sum_{i=1}^{C}\frac{T_i}{T_i+F_i}\label{eq:14}
\end{equation}

In the equations, $T$ represents the number of correctly classified samples,  $F$ denotes the number of misclassified samples, and $C$ represents the number of asteroid categories.

We built an ANN network based on the description in the article by Penttilä et al. (\citeyear{article19}), a one-dimensional data classification network (Transformer) based on the description in Weng et al.'s article (\citeyear{weng2021one}), and using classic machine learning algorithms and deep learning algorithms. We use the training dataset and validation dataset to train the network, and use the test dataset to test the network. We compared the classification results of these networks with those of the ASC-Net network we built, and the results are shown in Table \ref{tab:5}. Since the number of categories in our test dataset is balanced, the results for Acc and AP are are consistent in numerical values.
   \begin{table}[h]
      \caption{Test results on ASTD}
        \centering
         \begin{tabular}{ccc}
            \hline
            \noalign{\smallskip}
           Algorithms      & AP/\%&  Acc/\%\\
            \noalign{\smallskip}
            \hline
            \noalign{\smallskip}

            PCA+SVM& 84.50 & 84.50   \\
            Random Forest & 85.06  &    85.06      \\
            ANN         &90.86&90.86 \\
            ResNet-50   & 92.50 &92.50\\
            Transformer&91.67&91.67\\
           ASC-Net(Ours)&94.58&94.58\\
            \noalign{\smallskip}
            \hline
            \label{tab:5}
         \end{tabular}
   \end{table}

As shown in the experimental results, due to limitations such as poor non-linear fitting ability and limited high-dimensional feature extraction ability, the classification accuracy of traditional machine learning methods on the ASTD dataset is significantly lower than that of deep learning methods. In the realm of deep learning methods, the AP and ACC of the ANN, comprising only two fully connected layers, have already surpassed 90\% on the ASTD dataset. Meanwhile, the ResNet-50 network with a complex architecture shows further improvements in AP and ACC, reaching 92.50\% and 91.67\% respectively. This indicates that deeper network architectures indeed enhance classification accuracy, benefiting from their complexity. It is worth noting that ASC-Net consists of only nine hidden layers and an attention mechanism, demonstrating outstanding performance with an AP and Acc of 94.58\%. 
\begin{figure}[h]
    \centering
    \includegraphics[width=9cm]{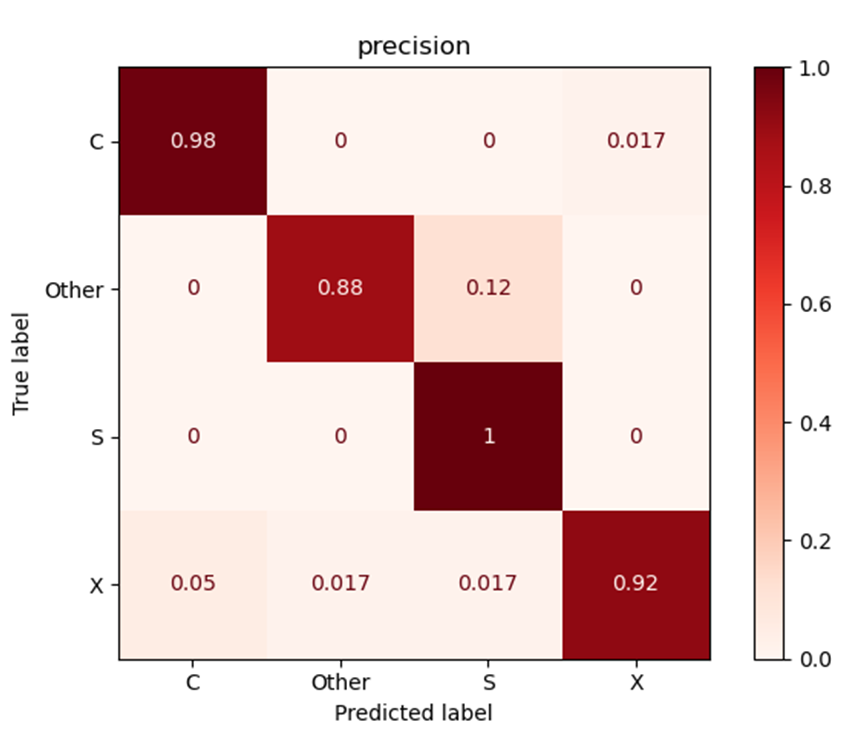}
    \begin{center}
    \caption{Confusion matrix of ASC-Net test results on ASTD}
    \label{fig:6}
    \end{center}
\end{figure}

The confusion matrix of ASC-Net's classification results on the ASTD dataset is depicted in Figure \ref{fig:6}. The horizontal axis represents the categories predicted by ASC-Net, while the vertical axis represents the actual categories of the asteroids corresponding to each spectrum. The main source of classification errors is the misclassification of other types of asteroids, with 12\% of other types of asteroids classified as S-type. 5\% of X-class asteroids should not be classified as C-class.

\subsubsection{Test on VisNIR}

To evaluate the classification performance of the ASC-Net classification network on asteroid spectra of different wavelength ranges, we conducted generalization tests on the publicly available asteroid spectral classification dataset VisNIR \citep{article19}. The spectra in the VisNIR dataset range from 0.45 {\textmu}m to 2.45 {\textmu}m and include both visible and near-infrared bands. For classification purposes, the simplified 11-classification method was used, which was derived from the Bus-DeMeo classification method. The comparison between the Bus-DeMeo classification method and the 11-classification method used in the VisNIR dataset is presented in Table \ref{tab:6}.
   \begin{table}[h]
      \caption{Comparison between the Bus-DeMeo taxonomy and the VisNIR taxonomy }
        \centering
         \begin{tabular}{cc}
            \hline
            \noalign{\smallskip}
           VisNIR\hspace{0.5em}taxonomy      & Bus-BDM\hspace{0.5em}taxonomy\\
            \noalign{\smallskip}
            \hline
            \noalign{\smallskip}

            A& A   \\
            B & B      \\
            C         &C\hspace{0.5em}Cb\hspace{0.5em}Cg\hspace{0.5em}Cgh\hspace{0.5em}Ch \\
            D   & D\\
            K&K\\
          L&L\\
          Q&Q\\
          S&S\hspace{0.5em}Sa\hspace{0.5em}Sv\hspace{0.5em}Sq\hspace{0.5em}Sr\\
          T&T\\
          V&V\\
          X&X\hspace{0.5em}Xc\hspace{0.5em}Xe\hspace{0.5em}Xk\\
            \noalign{\smallskip}
            \hline
            \label{tab:6}
         \end{tabular}
   \end{table}

In the 11-classification method used in the VisNIR dataset, the original C, Cb, Cg, Cgh, and Ch classes in the Bus-DeMeo classification method were merged into the C class, while the S, Sa, Sv, Sq, and Sr classes were merged into the S class. Similarly, the X, Xc, Xe, and Xk classes were merged into the X class. Additionally, the O and R classes with relatively small data sizes were removed.

VisNIR dataset includes various simulated datasets and raw datasets. We will use the simulated dataset as the training dataset and the raw dataset as the testing dataset. The training dataset includes 11 categories, with 200 entries per category for a total of 2200 entries. The testing dataset consists of 586 spectra of various asteroids.

The testing results of ASC-Net and other algorithms on the VisNIR dataset are presented in Table \ref{tab:7}. Due to the varying amounts of test data for each category in the VisNIR dataset, the mAP and Acc metrics are different. Therefore, both evaluation metrics can be combined to assess the network performance. 
\begin{table}[h]
      \caption{Test results on VisNIR dataset}
        \centering
         \begin{tabular}{ccc}
            \hline
            \noalign{\smallskip}
           Algorithms      & AP/\%&  Acc/\%\\
            \noalign{\smallskip}
            \hline
            \noalign{\smallskip}

            PCA+SVM& 56.13 & 57.26   \\
            Random\hspace{0.5em}Forest & 58.33  &   65.61      \\
            ANN         &91.37&90.68 \\
            ResNet-50   & 94.82 &92.15\\
            Transformer&97.03&91.07\\
           ASC-Net(Ours)&97.87&95.69\\
            \noalign{\smallskip}
            \hline
            \label{tab:7}
         \end{tabular}

   \end{table}

From the experimental results in Table \ref{tab:7}, it can be observed that even with the expansion of the spectral range of asteroid spectra to the near-infrared, ASC-Net still maintains a high average accuracy of 97.87\% and a high precision of 95.69\%. This indicates that ASC-Net exhibits good generalization performance, enabling high-precision classification of asteroid spectra inputs across different spectral ranges. Furthermore, compared to the results of the 4-class experiment on the ASTD dataset, although the number of classification categories in the VisNIR dataset is much larger than the 4 classes in ASTD, the classification accuracy of each algorithm has not significantly decreased. This suggests that the extension of the spectral range of asteroid spectra can provide more information to classification networks, thereby enhancing classification accuracy.
\begin{figure}[h]
    \centering
    \includegraphics[width=9cm]{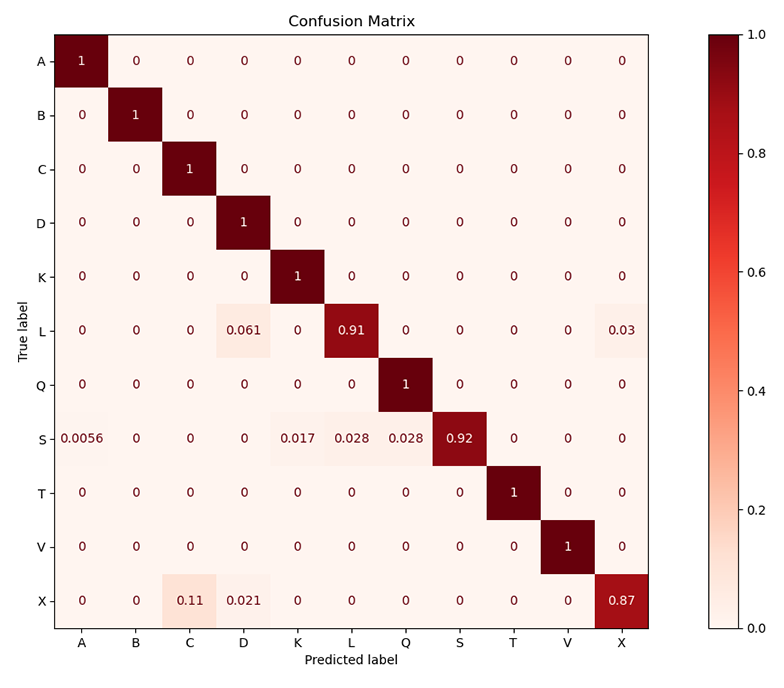}

    \caption{Confusion matrix of ASC-Net test results on VisNIR}
    \label{fig:7}

\end{figure}

The confusion matrix of ASC-Net's classification results on the VisNIR dataset is depicted in Figure \ref{fig:7}. On the VisNIR dataset, ASC-Net achieves a classification accuracy of 100\% for the majority of asteroid categories, with only a few spectral misclassifications observed in the S, X, and L classes.

\subsection{composition analysis Accuracy of AE-trans}
\label{subsec:tables}
\subsubsection{Test on ACAD}
To validate the performance of our asteroid spectral unmixing network, we also conduct comparative experiments on the unmixing effectiveness between the AE-trans and other machine learning and deep learning methods on our ACAD dataset. For the evaluation of the predicted spectral of endmembers output, we employ SAD (Spectral Angle Distance), which is an algorithm based on the overall similarity of spectral curves. It treats the spectrum of each pixel in an image as a high-dimensional vector and measures the similarity between spectra by calculating the angle between two vectors. The smaller the angle, the more similar the spectra, indicating a higher likelihood of belonging to the same class of material. The computation of SAD is depicted by Equation \ref{eq:15},  and  represent the actual spectrum and the predicted spectrum respectively.

\begin{equation}
SAD(\textbf{S},\textbf{\^{S}})=\frac{1}{R}\sum_{i=1}^{R}arccos 
  (\frac{<\textbf{s}_{(i)},\textbf{\^{s}}_{(i)}>}{||\textbf{s}_{(i)}||_2,||\textbf{\^{s}}_{(i)}||_2})\label{eq:15}
\end{equation}

In the formula, $\textbf{S}$ represents the slected material spectrum and $\textbf{\^{S}}$ represents the unmixed endmember spectrum. $R$ represents the number of endmembers. $i$ is the number of bands.

For assessing the abundance of material composition, we utilize Root Mean Square Error (RMSE) as an evaluation metric. RMSE is a commonly used measurement in statistics and machine learning to assess the average squared difference between actual and predicted values in regression problems. RMSE provides an indication of the average error between actual and predicted values. The calculation formula for RMSE is as follows:

\begin{equation}
RMSE(x,y)=\sqrt{\frac{1}{m}\sum_{i=1}^{m}(x_i-y_i)^2}
\end{equation}
In the equation, $x$ and $y$ represent the predicted and true values, and $m$ represents the number of spectra. A smaller RMSE indicates more accurate prediction results.

\begin{figure}[!htbp]
    \centering
    \includegraphics[width=9cm]{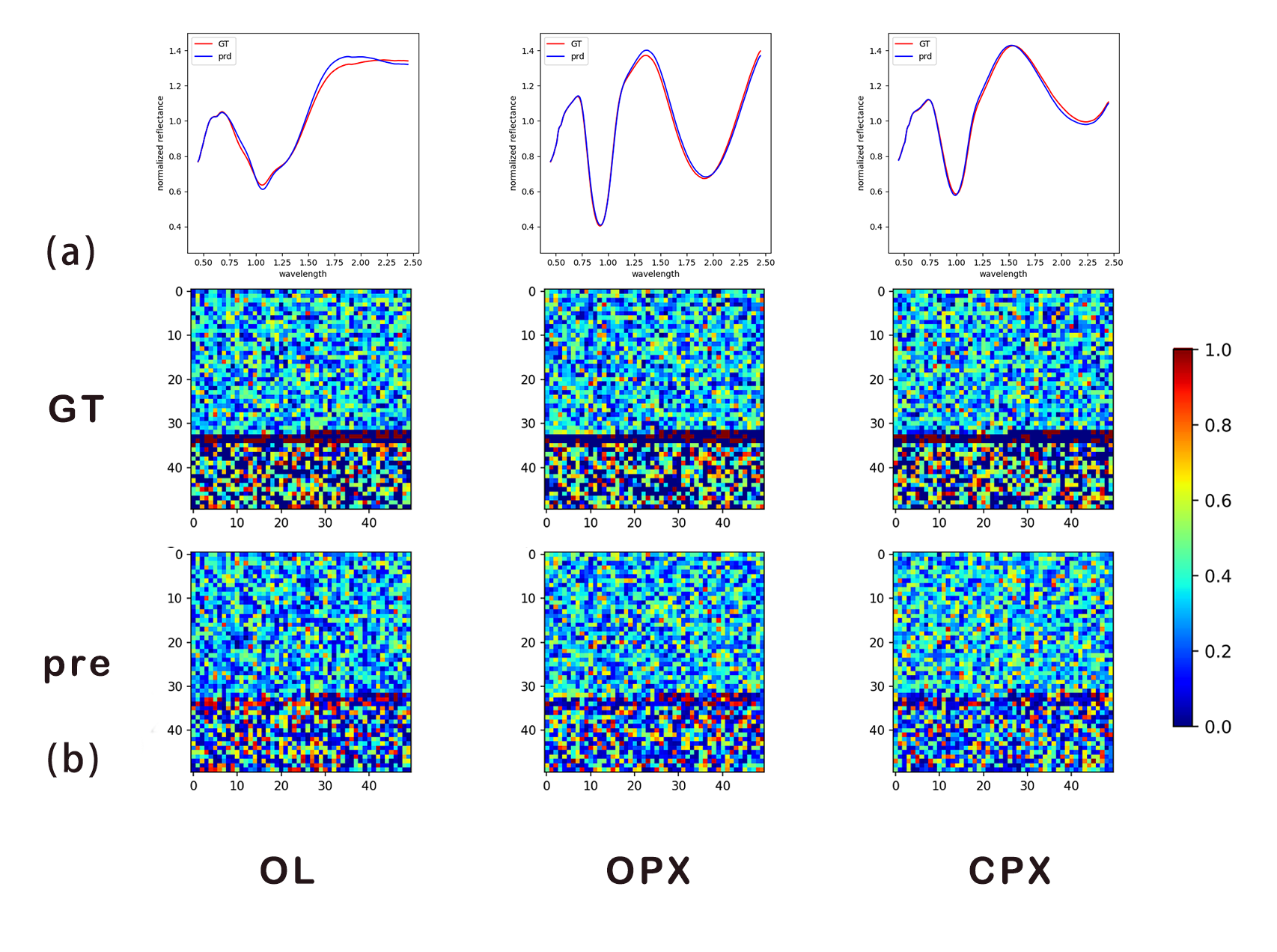}
    \caption{Confusion matrix of AE-tans test results on ACAD dataset:
    (a) : unmixed endmember spectra;
    (b): the first row is the true abundance map;
    the second row is Display of unmixed abundance map}
    \label{fig:ACAD}
\end{figure}

Our comparative experimental results are presented in Table \ref{tab:8}, the algorithms involved  include NMF (non-negative matrix factorization), PCA , deep learning spectral unmixing model HapkeCNN \citep{rasti2022hapkecnn}, and our AE-trans. 
\begin{table}[h]
      \caption{Test results of composition analysis}
         \centering
         \begin{tabular}{ccc}
            \hline
            \noalign{\smallskip}
           Algorithms      &SAD& RMSE\\
            \noalign{\smallskip}
            \hline
            \noalign{\smallskip}

            NMF& 0.1156 &0.2287   \\
             SparsePCA & 0.1968  & 0.2733\\
            HapkeCNN   &0.0635&0.3406 \\
            AE-trans (Ours)  & 0.0340 &0.1759\\

            \noalign{\smallskip}
            \hline
            \label{tab:8}
         \end{tabular}

   \end{table}
The experimental results demonstrate that our method has a prediction error of 0.0340 for end-member spectra, which provides a significant advantage compared to other methods. Based on the accurate prediction of end-member spectra, AE-trans also exhibits more precise abundance prediction for various components in the simulated mixture spectra, with a RMSE of only 0.1759, significantly lower than other methods.
The result is shown in Figure \ref{fig:ACAD}.

In section a of the figure, we compared three endmember spectra and material spectra, where the red GT represents the true value, which is the selected material spectrum, and the blue predicted represents the endmember spectrum extracted by the network. In section b of the figure, we convert the abundance matrix into the corresponding abundance map for display. The top row represents the ground-truth content map, and the bottom row represents the abundance map predicted by the network. The color change from blue to red represents the content from 0\% to 100\%.

\subsubsection{ Test on Korda D et al.'s unmixed dataset}
To test the generalization ability of our AAE-Net network, we used the dataset used by Korda D et al to train their network \citep{article22}. Their data includes not only three mineral modals, but also data on the different chemical composition contents of each modal. Therefore, the spectra corresponding to each mode may have some differences from the end member spectra we have selected.

\begin{figure}[h]
    \centering
    \includegraphics[width=9cm]{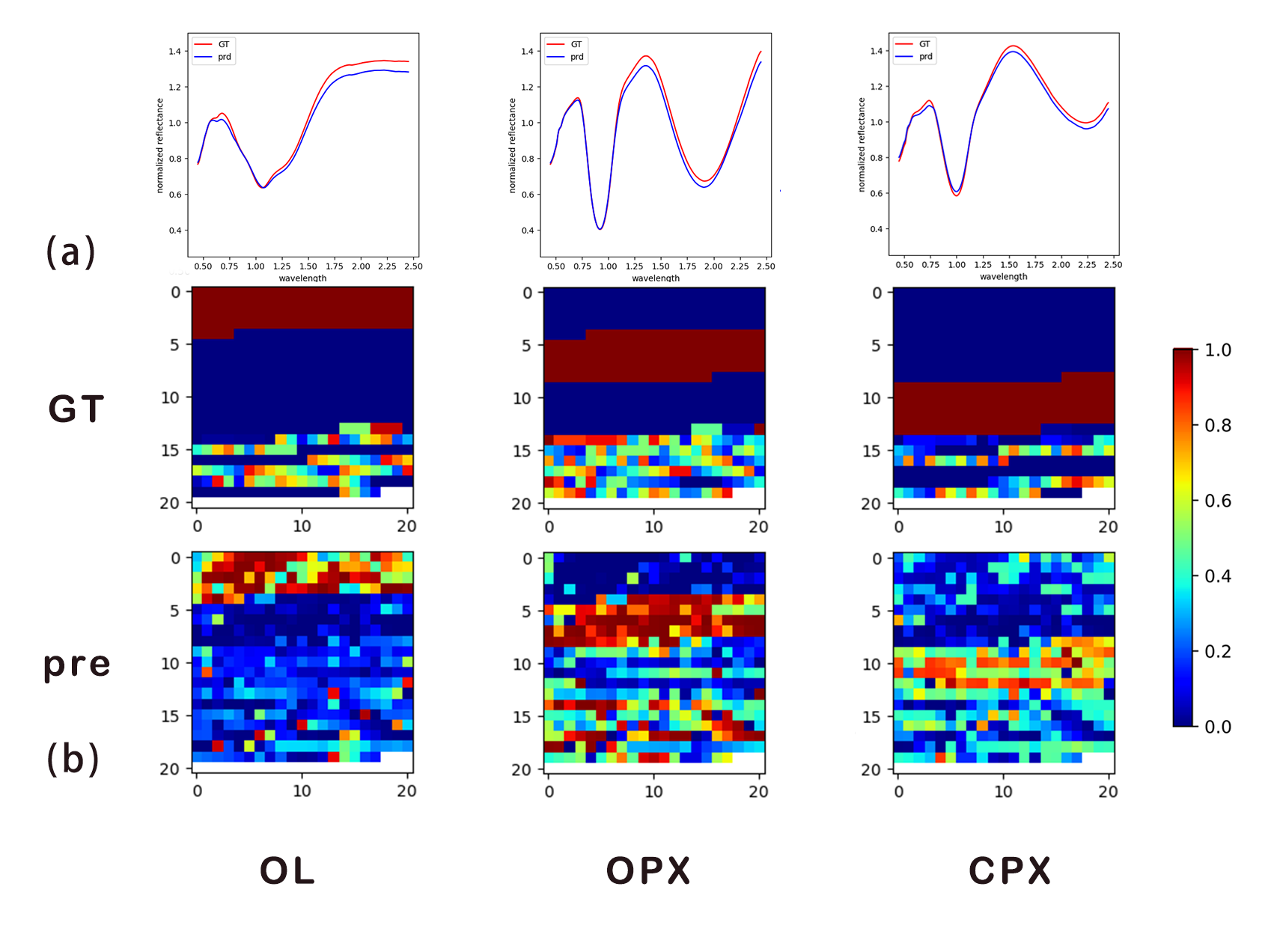}
    \caption{Confusion matrix of AE-trans test results on Korda D et al.'s unmixed dataset \citep{article22}:
    (a) : unmixed endmember spectra;
    (b): the first row is the true abundance map;
    the second row is Display of unmixed abundance map}
    \label{fig:8}
\end{figure}

The results obtained by inputting Korda D et al.'s unmixed dataset into our network obtain results with SAD=0.0334, RMSE=0.2330. The result is shown in the Figure \ref{fig:8}.

Using the same method to display the unmixing results, 417 spectra were converted into a 21 × 21 matrix in section b, where white pixels represent no data.

It can be seen that our unmixed abundance map is not significantly different from the actual abundance map, with the main difference being in the pure pixels (composed of only one mode). This is because our network unmixing takes the average spectrum of each mineral mode as a reference, and the network has not learned the influence of different chemical material compositions on the spectral shape of mineral modes. The analysis of different chemical material compositions can be a target for further research.

\subsection{Albedo Estimation of AAE-Net}
\label{subsec:tables}
\subsubsection{ Test on AAED}
To evaluate the predicted albedo of the network output, we used Mean Absolute Error (MAE) to measure the percentage error between the predicted and actual values, and we use the Mean Absolute Percentage Error (MAPE) to measure the percentage error between the predictions and the actual values.

\begin{equation}
MAE=\frac{1}{n} \sum_{i=1}^{n} \left | {\hat{y_{i}}- y_{i}}  \right | \label{eq:mae}
\end{equation}

\begin{equation}
MAPE = \frac{1}{n} \sum_{i = 1}^{n} \frac{\left | {\hat{y_{i}}- y_{i}}  \right |}{y_{i}} \times 100\%
 \label{eq:mape}
\end{equation}

$y_i$ and $\hat{y_{i}}$ respectively represent the actual albedo and predicted albedo, $n$ is the number of data.

The estimation results for AAED are shown as Table \ref{tab:AAED}. It can be seen that compared with the results of machine learning, the AAE-Net network has the smallest estimated albedo error and the best performance. At the same time, we assign the average albedo value corresponding to the type result given by the classification network ASC-Net. When comparing this result with that of the AAE-Net network, it can be observed that the albedo estimated by the AAE-Net network has smaller MAE and MAPE values, indicating a certain level of practicality. However, due to the normalization of current spectral data, there is a lack of information reflecting the absolute intensity of reflected light, which only reflects the compositional characteristics and taxonomy information of asteroids. Therefore, other relevant information may be introduced in the future to achieve better results.
\begin{table}[h]
      \caption{Test results on AAED }
         \centering
         \begin{tabular}{ccc}
            \hline
            \noalign{\smallskip}
           Algorithms      & MAE& MAPE(\%)\\
            \noalign{\smallskip}
            \hline
            \noalign{\smallskip}
            SVR& 0.0874 &59.20\\
            DT& 0.0678&51.80\\
            AdB &0.0695&59.79 \\
            Rid & 0.0639&58.71\\
            KRid&0.0654&53.46\\
            MLP&0.0743&91.61\\
            ASC-Net& 0.0652   &59.85  \\
           AAE-Net(Ours)&0.0441&40.35\\
            \noalign{\smallskip}
            \hline
            \label{tab:AAED}
         \end{tabular}
   \end{table}
We also classify asteroids by taxonomy and display the results as shown in the Table \ref{tab:10}. The network has relatively small errors in estimating the albedo of S-type and C-type asteroids, while X-type asteroids may have large errors in estimating albedo due to the small amount of data and large distribution range.
\begin{table}[h]
      \caption{Test results of different types of asteroids}
         \centering
         \begin{tabular}{ccc}
            \hline
            \noalign{\smallskip}
           Dataset      & MAE&  MAPE(\%)\\
            \noalign{\smallskip}
            \hline
            \noalign{\smallskip}

           S& 0.0410 & 28.86  \\
            C & 0.0161  &  43.56      \\
            X  &0.1584&86.47  \\

            \noalign{\smallskip}
            \hline
 
         \end{tabular}
      \label{tab:10}
   \end{table}

It can be seen that the results of X-type are much worse than those of S-type and C-type. Perhaps the X-type itself is composed of several types with significant dif-979 strong reflectivity, such as the E-type with an average reflectivity of 0.57 and the M-type with an average reflectivity of 0.14, which may require more detailed classification. Therefore, we are currently only focusing on S-type and C-type asteroids with better results.

\subsection{Application} \label{sec:cite}

\subsubsection{ Testing ASC-Net with Actual Data}

For our constructed platform for asteroid spectral classification and compositional analysis, we are conducting preliminary applications using existing small asteroid spectral data. We select spectral inputs from several asteroids to test our platform.
We selected two asteroids from each of the three types of asteroids: S-type, C-type, and X-type as test data. These asteroid data were not included in the previous training and testing datasets, and it is best to select asteroids that have undergone close range exploration missions. 
The tested asteroids include Ceres (1), Bennu (101955), Itokawa (25143), Eros (433), Kalliope (22), and Angelina (64). The small asteroid spectral data used in the tests are sourced from the VisNIR dataset provided by Penttilä \citep{article19}. The classification results are shown in Table \ref{tab:11}.
\begin{table}[h]
      \caption{Spectral taxonomy results of 6 asteroids}
          \centering
         \begin{tabular}{cccc}
            \hline
            \noalign{\smallskip}
                 &Type& The \hspace{0.5em}first&The\hspace{0.5em} second\\
            \noalign{\smallskip}
            \hline
            \noalign{\smallskip}

            Ceres& C &C: 89.6\% &X: 4.4\%   \\
            Bennu & C  &C: 88.7\%&S: 4.6\%\\
            Itokawa   &S&S: 87.2\% &Other: 10.9\% \\
            Eros  & S &S: 84.1\% &Other: 14.0\%\\
            Kalliope   &X&X: 88.8\%& C: 6.2\%\\
            Angelina  & X &X: 90.9\% &C: 5.0\%\\
            \noalign{\smallskip}
            \hline
            \label{tab:11}
         \end{tabular}

   \end{table}

\begin{figure*}[!htbp]
    \centering
    \includegraphics[width=16cm]{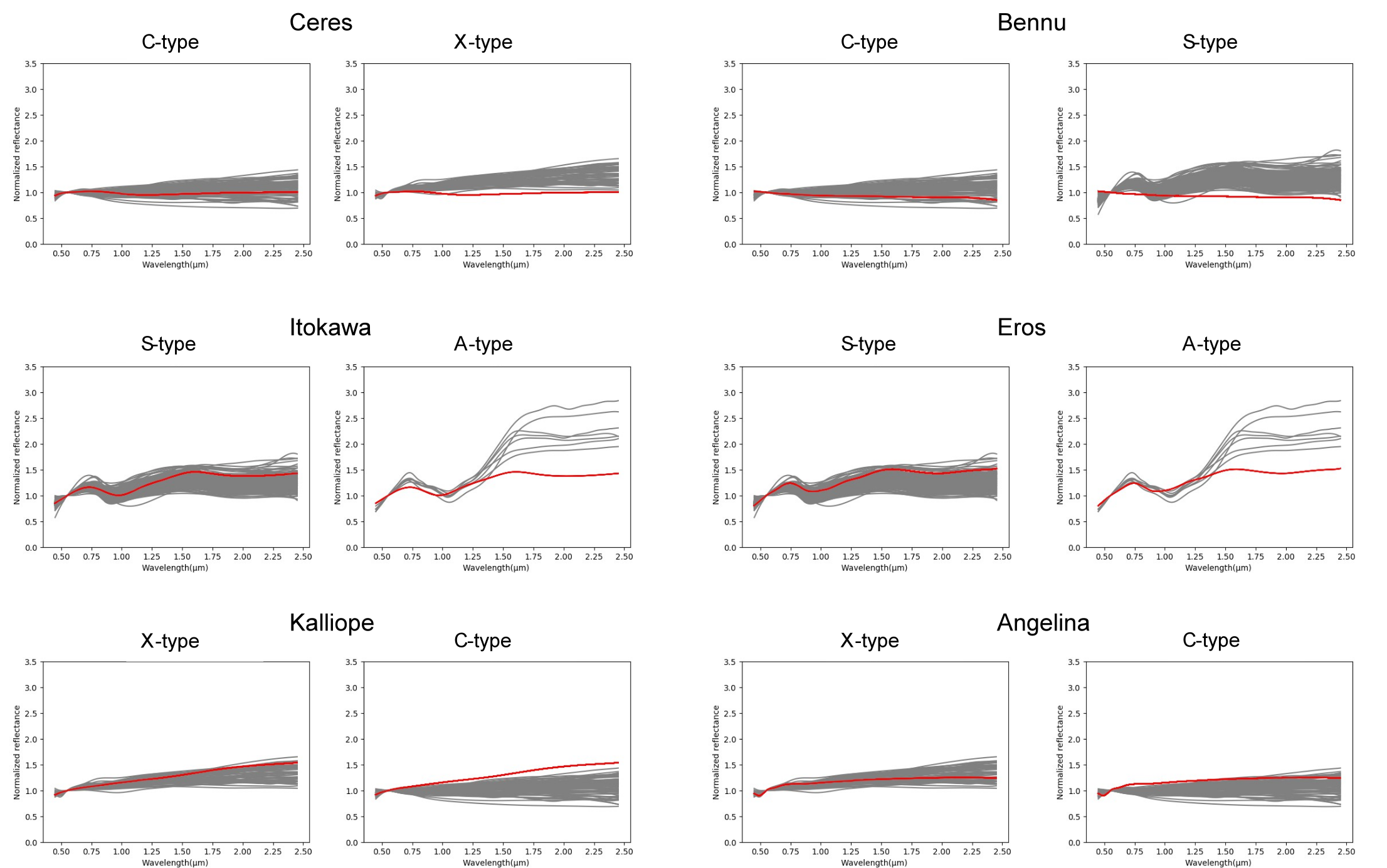}

    \caption{Comparison of the spectra of tested asteroids to the spectra of the simulated samples in their predicted classes.}
    \label{fig:10}

\end{figure*}

In Table \ref{tab:11}, we provide the top two highest probability predictions of ASC Net for classifying these six asteroids. Figure \ref{fig:10} shows a comparison of the spectra of these six asteroids with the top two predicted categories with the highest probabilities. The likelihood of these six asteroids being classified into the highest probability category is much higher than the likelihood of being classified into the second category, and they are consistent with the currently assigned category. For Ceres and Bennu, their overall spectra are relatively flat and have low reflectance, leading the network to perceive a certain degree of similarity with X-type and S-type asteroids. Similarly, Kalliope and Angelina are also considered to have a possibility of being C-type asteroids for the same reasons. For Itokawa and Eros, we observed a high probability of being classified as the "other" taxonomy. Therefore, we conducted further sub-division and found that they are likely to belong to the A sub-class, with probabilities of 7.32\% and 9.11\%, respectively. From Figure \ref{fig:10}, it can be seen that this is possibly due to the spectral similarities between Itokawa and Eros with the A class to some extent. However, in the near-infrared spectrum, their reflectance is lower, aligning more with the spectral characteristics of S-type asteroids. Therefore, ASC-Net still indicates a probability of over 80\% for them being S-type asteroids.

\subsubsection{ Testing AE-trans with Actual Data}

We used the Eros (433) and  Itokawa (25143) spectral data collected by Korda D et al \citep{article23} and applied them to our S-type unmixing network. The dataset is shown in the following Figure \ref{fig:as_data}.

\begin{figure}[!htbp]
    \centering
    \includegraphics[width=9cm]{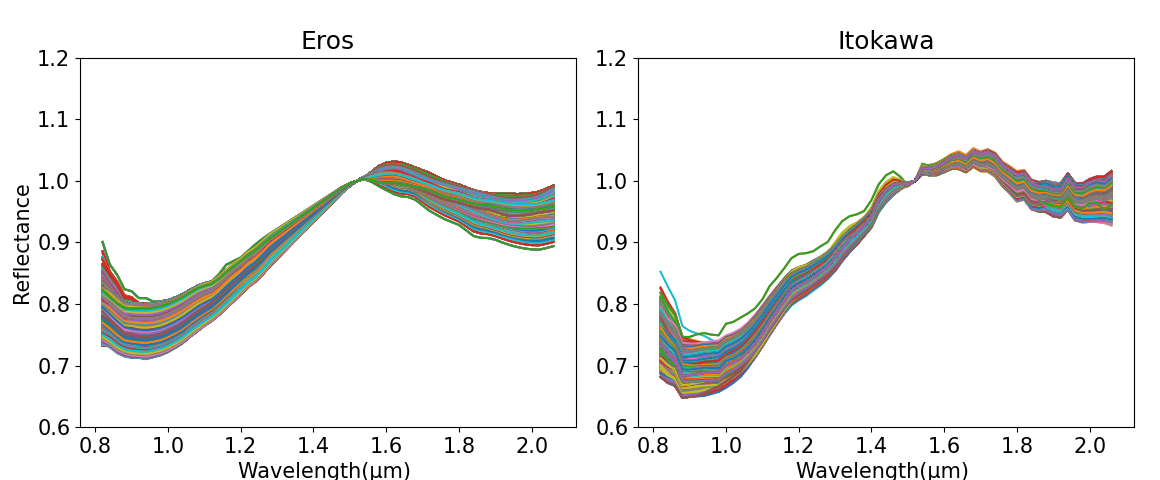}
    \caption{Spectral data of Eros (left) and Itokawa (right) }
    \label{fig:as_data}
\end{figure}

The corresponding results obtained were compared with those presented in Korda D's paper.The results are shown in Table \ref{tab:12}.

\begin{table}[h]
      \caption{Unmixing of Eros spectral data and Itokawa spectral data and  laboratory composition analysis.  }
      \fontsize{9pt}{12pt}\selectfont
          \centering
         \begin{tabular}{ccccc}
            \hline
            \noalign{\smallskip}
         &\multicolumn{1}{c}{} &\multicolumn{1}{c}{Asteroid-spectra} & \multicolumn{1}{c}{AE-trans} &  \multicolumn{1}{c}{Lab$^{\ast}$}\\
         &&\citep{article23}&(ours)&\\
            \noalign{\smallskip}
            \hline
            \noalign{\smallskip}

           &OL(\%)&$57.6\pm4.5$&$42.6\pm8.0$&	\\
    Eros&OPX(\%)&$28.9\pm1.8$&	$37.2\pm6.1$&\\
          &CPX(\%)&$13.5\pm4.1$& $20.2\pm8.2$&\\
\hline
 &OL(\%)&$69.0\pm1.5$&$59.9\pm1.8$&	75.2\\
Itokawa&OPX(\%)&$27.0\pm1.7$& $37.1\pm1.5$&21.5\\
&CPX(\%)&$4.0\pm1.9$& $3.0\pm1.6$&3.3\\

            \hline
            \noalign{\smallskip}
         \end{tabular}  

\begin{tablenotes}
      \item[a]  \textbf{Note}. Itokawa particle analyses were adopted from \citep{nakamura2014mineral} and \citep{tsuchiyama2014three}. Modal mineral compositions are always normalised to OL + OPX + CPX = 100 vol\% by \citep{article23}.
    \end{tablenotes}
   \label{tab:12}
   \end{table}

Our network unmixing results are generally consistent with the network results and laboratory analysis results of Korda D et al., with Itokawa asteroid having the highest OL content, followed by OPX content, and CPX content being the lowest. However, the OL content in our network is relatively low, while the OPX content is relatively high.

\subsubsection{ Testing AAE-Net with Actual Data}

After obtaining the asteroid types through the classification network, we input the spectral data of the S-type and C-type asteroids that we are interested in into the corresponding albedo estimation network. The obtained albedo is compared with the collected albedo in the dataset, as shown in Figure \ref{fig:albedo_E} and Table \ref{tab:13}:

\begin{figure}[!htbp]
    \centering
    \includegraphics[width=9cm]{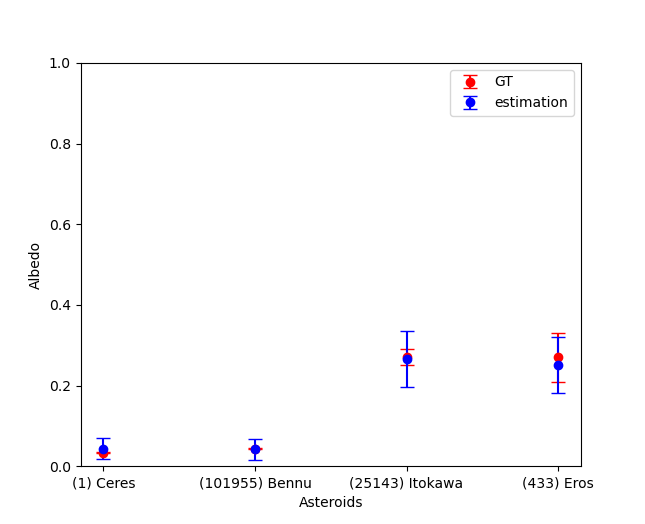}
    \begin{center}
    \caption{Compare the predicted albedo of asteroids with the albedo in the dataset. The blue dot represents the predicted albedo, and the red dot represents the dataset albedo}
    \label{fig:albedo_E}
    \end{center}
\end{figure}

\begin{table}[h]
      \caption{Estimation of albedo for 4 asteroids}
          \centering
         \begin{tabular}{cccc}
            \hline
            \noalign{\smallskip}
          Type& Asteroid      & Albedo&	Estimated\hspace{0.5em} Albedo\\
            \noalign{\smallskip}
            \hline
            \noalign{\smallskip}

          C &Ceres	&$0.034\pm 0.001$ &	$0.044\pm 0.026$\\
          C& Bennu	&$0.044 \pm 0.002$ &	$0.043\pm 0.026	$	 \\
           S &Itokawa&	$0.270 \pm 0.020$& 	$0.266\pm 0.069$	\\
           S & Eros	&$0.270 \pm 0.060$ 	&$0.251\pm 0.069$	\\

            \noalign{\smallskip}
            \hline
 
         \end{tabular}
      \label{tab:13}
   \end{table}

It can be seen that the prediction results of these asteroids are relatively good.

2016HO3 (469219) belongs to an S-type asteroid, so the results related to S-type asteroids are of great concern to us. The network provides the best estimation of albedo for S-type asteroids, with the minimum MAE and the minimum MAPE.

\section{Conclusions} \label{sec:cite}

By introducing advanced deep learning transformation networks with multi head attention mechanisms, we have developed a platform that provides a high-precision method for asteroid spectral classification, albedo estimation, and composition analysis. Through this platform, spectral analysis support can be provided for China's Tianwen-2 mission and even broader deep space exploration missions. Our platform consists of three neural networks: ASC-Net for spectral classification, AAE-Net for albedo estimation and AE-trans for composition analysis.

For the ASC-Net spectral classification network, The test results show that the 4-class accuracy of ASC-Net on the SMASS II database reaches 94.58\% and the 11-class accuracy on the VisNIR dataset reaches 95.69\%.
 For the AE-trans compositional analysis network, we trained and tested the network using a simulated dataset based on material spectra from the RELAB database. The network predict a material spectral angular distance of only 0.0340 and a root mean square error of 0.1759 for end-member abundance. The unmixing results in another dataset by Korda D et al. 
For the AAE Net albedo estimation network, the average absolute error in estimating the albedo of S-type asteroids is 0.0410, and the relative percentage error is 28.86\%.

Based on the neural networks and platform we have constructed, more in-depth research can be conducted in the future. Subsequent studies can explore a wider range of spectral wavelengths and sampling intervals for asteroid spectra to test the accuracy and generalization of the ASC-Net spectral classification network, and add more spectral data to ensure the balance of various types of data. For the AE-trans composition analysis network, we can try adding more endmembers or further subdividing various chemical compositions under different mineral modes, so that AE-trans can analyze the composition of asteroids more comprehensively. Search for unnormalized spectral data to preserve sufficient original features for input into the AAE-Net albedo estimation network, or provide more asteroid feature values to the network.

\begin{acknowledgments}
This research was funded by \emph{National Science and Technology Major Project (2022ZD0117401)} and \emph{the National Defense Science and Technology Innovation Special Zone Project Foundation of China grant number 19-163-21-TS-001-067-01.}
       We would like to kindly acknowledge  A. Penttilä for providing us with the VisNIR asteroid datasets, utilized in this study
\end{acknowledgments}



\bibliographystyle{aasjournal}



\end{document}